\documentclass[twocolumn]{aastex631}

\usepackage{amsmath}
\usepackage{amsfonts}
\usepackage{graphicx}
\usepackage{aas_macros}
\usepackage{mathtools}

\definecolor{mpl_blue}{HTML}{1F77B4}
\definecolor{mpl_orange}{HTML}{FF7F0E}
\definecolor{mpl_green}{HTML}{2CA02C}
\definecolor{mpl_red}{HTML}{D62728}

\DeclarePairedDelimiterX{\infdivx}[2]{(}{)}{%
  #1\;\delimsize\|\;#2%
}
\newcommand{\kldiv}{D_{\mathrm{KL}}\infdivx}

\newcommand{\code}[1]{\texttt{#1}}

\begin{document}
\title{Gravitational Wave Statistics for Pulsar Timing Arrays: Examining Bias from Using a Finite Number of Pulsars}
\author{Aaron D.~Johnson} 
\affiliation{Center for Gravitation, Cosmology and Astrophysics, University of Wisconsin--Milwaukee, P.O. Box 413, Milwaukee WI, 53201, USA}

\author{Sarah J.~Vigeland} 
\affiliation{Center for Gravitation, Cosmology and Astrophysics, University of Wisconsin--Milwaukee, P.O. Box 413, Milwaukee WI, 53201, USA}

\author{Xavier Siemens} 
\affiliation{Department of Physics, Oregon State University, Corvallis, OR 97331, USA}
\affiliation{Center for Gravitation, Cosmology and Astrophysics, University of Wisconsin--Milwaukee, P.O. Box 413, Milwaukee WI, 53201, USA}

\author{Stephen R.~Taylor} 
\affiliation{Department of Physics and Astronomy, Vanderbilt University, 2301 Vanderbilt Place, Nashville, TN 37235, USA}

\begin{abstract}
Recently, many different pulsar timing array (PTA) collaborations have reported strong evidence for a common stochastic process in their data sets. 
The reported amplitudes are in tension with previously computed upper limits.
In this paper, we investigate how using a subset of a set of pulsars biases Bayesian upper limit recovery. 
We generate 500 simulated PTA data sets based on the NANOGrav 11-year data set with an injected stochastic gravitational wave background (GWB).
We then compute upper limits by sampling individual pulsar likelihoods, and combine them through a factorized version of the PTA likelihood to obtain upper limits on the GWB amplitude using different numbers of pulsars.
We find that it is possible to recover an upper limit (95\% credible interval) \textit{below} the injected value, and that it is significantly more likely for this to occur when using a subset of pulsars to compute the upper limit. 
When picking pulsars to induce the maximum possible bias, we find that the 95\% Bayesian upper limit recovered is below the injected value in 10.6\% (53 of 500) of realizations.
Further, we find that if we choose a subset of pulsars in order to obtain a lower upper limit than when using the full set of pulsars, the distribution of upper limits obtained from these 500 realizations is shifted to lower amplitude values.

\end{abstract}

\section{Introduction}
Pulsar timing arrays (PTAs) aim to detect gravitational waves in the nanohertz frequency regime by looking for correlations between times of arrival of radio signals from millisecond pulsars (MSPs;  \citealt{taylor_nanohertz_2021}). 
Pulsar timing models predict the pulse times of arrival from a pulsar based on that pulsar's astrophysical properties. The differences between the predicted and measured times of arrival are the timing residuals \citep{verbiest2021pulsar}.
By modeling these residuals, we attempt to reveal gravitational wave signals hidden in our data.

The first such gravitational wave signal detected by PTAs is expected to come from a stochastic gravitational wave background (GWB) made up of gravitational waves emitted by a cosmological population of supermassive binary black holes (SMBBH; \citealt{rosado2015expected}). 
Assuming that these SMBBHs are circular and evolve only due to gravitational wave emission, the characteristic strain spectrum is given by \citep{phinney_practical_2001}
\begin{equation}
    h_c(f) = A_\text{GWB}\left( \frac{f}{f_\text{yr}} \right)^{-2/3},
\end{equation}
where the amplitude $A_\text{GWB}$ depends on the SMBBH population and galaxy merger rate, 
and $f_\text{yr}$ is the reference frequency corresponding to $1\;\text{yr}^{-1}$. 
Based on models of the supermassive binary black hole population, 
we expect to detect the GWB with PTAs within the next 5 years \citep{taylor_are_2016, pol_astrophysics_2021}. 

Recently, multiple PTAs have found evidence of a 
common spectrum stochastic process. 
The NANOGrav collaboration reported a common spectrum process with 
median strain amplitude $1.92 \times 10^{-15}$ in an analysis of their 
12.5-year data set \citep{NANOGrav_12p5yr_gwb}, 
the PPTA reported a median amplitude $2.2 \times 10^{-15}$ \citep{goncharov_evidence_2021}, the EPTA reported a median amplitude of $2.95 \times 10^{-15}$ \citep{epta_detection}, and the IPTA which used combined data from constituent collaborations' older data sets reported a median amplitude of $2.8 \times 10^{-15}$ \citep{ipta_dr2_gwb}.
The amplitude of this process is in tension with 
some previously published upper limits
on the amplitude of the GWB.

The NANOGrav collaboration placed an upper limit of $A < 1.45 \times 10^{-15}$ based on an analysis of 34 pulsars timed for up to 11 years \citep{NANOGrav_11yr_gwb}.
These pulsars consist of the ones from the 11-year data set that have been timed for more than 3 years.
The PPTA placed an upper limit of $A < 10^{-15}$ based on an analysis of four pulsars that had the highest timing precision and were timed for up to 11 years \citep{srl+2015}.
The EPTA placed an upper limit of $A < 3 \times 10^{-15}$ based on an analysis of six pulsars timed for up to 18 years \citep{lentati_european_2015}.
These six pulsars were chosen to minimize the dimensionality required to search over.
Additionally, the least sensitive pulsar of the six affected the result at the 2\% level.

There are several possible explanations for this apparent discrepancy 
between earlier results and the most recent ones. 
Early work did not model uncertainty in the position of the 
Solar System barycenter, and as shown in \citet{vts+2020}, 
the choice of Solar System ephemeris can significantly 
affect detection statistics. 
The choice of prior on pulsar intrinisic red noise also 
has a significant effect 
on the upper limit on a common stochastic process, 
as shown in \cite{hss+2020}, 
due to covariance between the two.

Here we investigate how Bayesian GWB upper limits are 
affected by the use of a finite number of pulsars. 
We generate simulated PTA data with an injected GWB, 
and compare the Bayesian upper limits computed by analyzing the entire PTA 
versus using only a subset of the pulsars. 
We show that a wide range of possible upper limits 
can be computed when only a small number of pulsars 
are used to compute the upper limit. 
Furthermore, it is possible to find an upper limit that is \textit{lower} than the injected value of the GWB, and this occurs more often when using a subset of pulsars.

This paper is organized as follows.
In Section \ref{sec:methods}, we discuss the procedure by which we simulate pulsars and compute upper limits.
Section \ref{sec:results} details how the upper limits change with different combinations of pulsars.
Finally, in Section \ref{sec:conclusion}, we discuss our results and make concluding remarks about what this means for the number of pulsars that are used in pulsar timing array data sets.

\section{Methods}
\label{sec:methods}
We use methods which are to a large extent the same as previous papers that set Bayesian upper limits \citep{nanograv_9yr_gwb, NANOGrav_11yr_gwb, lentati_european_2015, srl+2015}.
The significant differences include using a factorized PTA likelihood and grid approximating the posterior for each individual pulsar instead of sampling using a Metropolis-Hastings MCMC algorithm.
All models here, as in previous Bayesian upper limit papers, use a 30 frequency power law pulsar intrinsic red noise and GWB given by
\begin{equation}
    \rho(f) = \frac{A^2}{12\pi^2}\frac{1}{T}\left(\frac{f}{\text{yr}^{-1}}\right)^{-\gamma}\text{yr}^2,
\end{equation}
where $\rho(f) = S(f) \Delta f$ where $S(f)$ is the power spectral density and $\Delta f = 1 / T$.

All of the results in this paper use Bayesian methods.
The frequentist methods which have been used previously to set upper limits via the optimal statistic \citep{anholm2009optimal, demorest2012limits, chamberlin2015time} are not considered here.
Importantly, the Bayesian and frequentist upper limits have different interpretations and do not coincide in general \citep{rover2011bayesian}. 
Furthermore, the optimal statistic, which was used to set upper limits in \cite{nanograv_9yr_gwb}, looks at only the cross-correlations between different pulsars, 
while the Bayesian methods used to set upper limits in previous papers look only at 
auto-correlations, so the two are fundamentally different and it is difficult to compare them.

\subsection{Factorized likelihood}
\label{sec:factorizedlike}
When inter-pulsar correlations are not included, the PTA likelihood can be factored into a product of individual pulsars \citep{NANOGrav_12p5yr_gwb, taylor_parallelized_2022},
\begin{equation}
    p\left(\left\{d_{j}\right\}_{N} \mid\{\vec{\theta}_{j}\}_{N}, A_{\mathrm{GWB}}\right)=\prod_{j=1}^{N} p\left(d_{j} \mid \vec{\theta}_{j}, A_{\mathrm{GWB}}\right),
\end{equation}
where $d_j$ are the data, $\theta_j$ are the intrinsic noise parameters, and $A_{\mathrm{GWB}}$ is the common red process amplitude for the $j$th pulsar. Using the factorized likelihood allows for rapid computation of upper limits with more than one pulsar.
We use a grid approximation on the individual pulsar models as described in Section \ref{sec:implementation}, then multiply the marginalized common red process amplitude posteriors for each pulsar that we want in the combined upper limit.
These new posteriors are then reweighted from a log-uniform to a uniform prior on $A_\text{GWB}$ by multiplying by
\begin{equation}
    f(x) = \frac{10^x}{10^{x_\text{max}} - 10^{x_\text{min}}},
\end{equation}
where $x = \log_{10}A_\text{GWB}$, and $x_\text{max}$, $x_\text{min}$ are the maximum and minimum values of the uniform prior for the log amplitude.
From this reweighted marginalized posterior, we can take the 95\% Bayesian upper limit easily by interpolating the posterior and using a cumulative sum until we reach 0.95,
\begin{equation}
    \sum_{x_\text{min}}^{x_\text{95\%}} p(A_\text{GWB}\mid d) = 0.95,
\end{equation}
and then finding the $\log_{10}A_\text{GWB}$ value corresponding to $x_\text{95\%}$ where the sum was truncated.
All following discussions use the 95\% upper limit as computed here.

\subsection{Simulations}
We simulate 500 sets of pulsars using \code{TEMPO2} \citep{edwards_tempo2_2, hobbs_tempo2_1} and \code{libstempo} \citep{vallisneri_libstempo_2020}, 
with the observation baselines, observing cadences, and noise properties based on the 11-year NANOGrav data set \citep{NANOGrav_11yr_dataset}.
The full 11-year NANOGrav data set contains 45 pulsars.
Due to the large number of upper limits that need to be computed for following sections, we only simulate 22 of the 45 pulsars that have been timed for more than six years.
Pulsars with shorter timing baselines contribute less to the upper limit than ones that have been observed for many years.
Because of this, we do not expect that removing these pulsars will significantly affect the results here.

Each pulsar contains white noise related to the uncertainty in the pulsar times of arrival (EFAC $=1$), intrinsic red noise similar to that in the 11 year data set, and an injected GWB with an amplitude $A_\text{GWB} = 10^{-15}$ and a spectral index, $\gamma_\text{GWB} = 13/3$.
The GWB injection includes Hellings and Downs cross-correlations \citep{hellings_upper_1983}, but we only use the auto-correlations to set the upper limits, as was done in many previous PTA papers \citep{srl+2015, nanograv_9yr_gwb, NANOGrav_11yr_gwb, lentati2016spin}.
\cite{lentati_european_2015} did use cross-correlations, but found that the upper limits were consistent with their auto-correlation only analysis.
Similarly, we also find that including cross-correlations does not change the upper limit in the cases that the upper limit falls below the injected amplitude.
The auto-correlations dominate the recovery of a GWB when the number of pulsars $N_{p}$ is relatively small since all of the cross-correlation coefficients are less than 1; however, for large numbers of pulsars, the cross-correlations become more significant since the number of cross-correlation terms increases as $\mathcal{O}(N_{p}^2)$.
In this particular set of simulated pulsars, the cross-correlations are too weak compared to the auto-correlations to affect the upper limits.

\subsection{Software and implementation}
\label{sec:implementation}
We use \code{enterprise} \citep{ellis_justin_a_enterprise_2020} to set up a model for our simulated pulsar sets with priors as in Table \ref{tab:model}.
Here we use a grid approximation to obtain each pulsar's posterior which is rendered effective by the low dimensionality of the parameter space.
We use a power law model for both the red noise intrinsic to each pulsar and the red noise common among all the pulsars.
Therefore, we set up our grid for each individual pulsar over the (1) intrinsic red noise amplitude, (2) intrinsic red noise spectral index, and (3) the common red process amplitude.
Care must be taken since models from \code{enterprise} return a log-likelihood.
To facilitate the use of the grid approximation, we subtract the maximum log-likelihood evaluated on the grid from all points before exponentiation to return the posterior which could then be marginalized.
We use a Nelder-Mead algorithm \citep{gao_implementing_2012, 2020SciPy-NMeth} to find this maximum starting from several random locations in the parameter space.
Once the maximum was found, we evaluate 300 points of the $\log_{10}A_{\text{GWB}}$ marginalized posterior.
To evaluate each point, we marginalize over the intrinsic red noise parameters simultaneously using \code{scipy.integrate.dblquad} \citep{2020SciPy-NMeth}.
This reduces the number of evaluations by allowing the adaptive integration routine to decide how many points are required instead of using a uniform grid.
We model each pulsar individually and then post-process using the factorized likelihood as discussed in Section \ref{sec:factorizedlike}.

\begin{table}[htp]
    \begin{tabular}{c|c|c}
        Parameter & Description & Interval\\
        \hline
        $A_{\text{RN}}$  & log-uniform & $[-20, -11]$ \\
        $\gamma_{\text{RN}}$ & uniform & $[0, 7]$ \\
        $A_{\text{GWB}}$ & log-uniform & $[-20, -11]$ \\
        $\gamma_{\text{GWB}}$& constant & $13/3$ \\
        $\text{EFAC}$ & constant & $1$\\
    \end{tabular}
    \caption{Priors for the model used to analyze each individual simulated pulsar. Intrinsic red noise parameters have been labeled with ``RN", and the common red process amplitude and spectral index have been labelled with ``GWB." The spectral index for the common red process has been fixed in each model to 13/3. EFAC is a multiplicative factor on the times of arrival uncertainties.}
    \label{tab:model}
\end{table}

\section{Results}
Using the above methods and software, we investigate how bias may appear when using a subset of pulsars.
We start this section with three specific combinations and how their cumulative upper limits change as we add more pulsars.
Next, we generalize to all possible combinations and average over all realizations to discover trends in how adding more pulsars affects the distribution of upper limits that are possible.
We then investigate a single combination for every realization to see how upper limits change as we add more pulsars.
Some of these pulsars hold more influence over the upper limits than others.
By using the Kullback-Leibler divergence \citep{kullback_information_1951}, we enumerate and examine these pulsars.
Finally, we investigate bias by comparing the distributions of upper limits obtained when computing the minimum cumulative upper limit given by a sequence of 22 pulsars for all 500 realizations.

\label{sec:results}

\subsection{Combinations and upper limits}
\label{sec:orderings}
One of the goals of this work is to investigate how the choice of which pulsars to use affects the computation of upper limits. 
Initially, we consider three different combinations: time-span, single-pulsar upper limit, and the combination which uses a greedy algorithm to get a low upper limit.
\begin{enumerate}
    \item The time span combination is, as its name suggests, a combination which sorts pulsars by their observation time span.
    \item After taking each pulsar and individually computing an upper limit, we can order these upper limits from lowest to highest. We call this the single-pulsar upper limit combination. This was the combination used in the NANOGrav 9-year stochastic gravitational wave background search \citep{nanograv_9yr_gwb}. When using this method, the upper limit dropped to a minimum and then increased with each added pulsar until it eventually saturated.
    \item The last combination is a greedy algorithm in which we build up the upper limit pulsar by pulsar. The lowest individual pulsar upper limit takes the first slot in the combination. Next, the upper limit is computed for the first slot and each of the remaining 21 pulsars. The pulsar from the remaining 21 that gives the lowest two-pulsar upper limit is put into the second slot. By continuing in this fashion until all pulsars have been used, we find the combination attains a minimum which is much lower than the time span combination.
\end{enumerate}

\begin{figure}[h]
\centering
\includegraphics[width=\columnwidth]{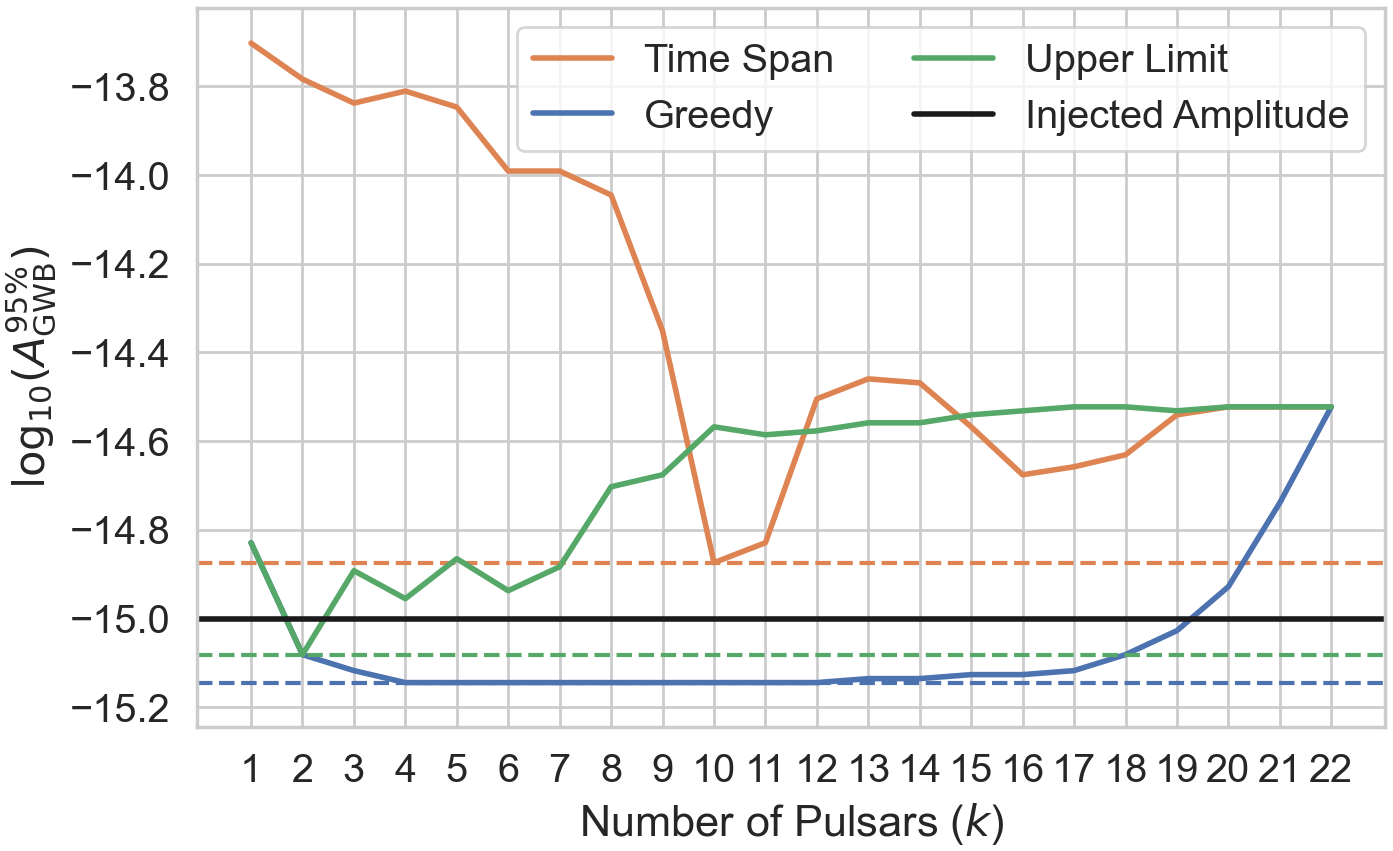}
\caption{Cumulative upper limits computed using three different upper limit combinations. While the different combinations vary significantly, they agree once the last pulsar has been added. The dashed lines are the minimum values that are achieved by each combination. Both the single-pulsar upper limit and the greedy upper limit combinations drop below the injected value when a subset of the pulsars are used.}
\label{fig:uls}
\end{figure}

Because there is only one combination when using all 22 pulsars, the combinations' upper limits converge as we use more pulsars.
However, these three combinations clearly show that there can be a large variance in upper limits when we use fewer pulsars.
One (particularly bad) realization using the three combinations discussed here is shown in Figure~\ref{fig:uls}.
Stopping with too few pulsars in either the single-pulsar upper limit or greedy upper limit combination returns a value that is below the injected amplitude.
In the greedy upper limit combination this is especially pronounced: the upper limits calculated with between 2 and 19 pulsars yield a value below the injected amplitude.

\subsection{All combinations}
\label{sec:single_combos}

\begin{figure*}[b]
\centering
\includegraphics[width=\textwidth]{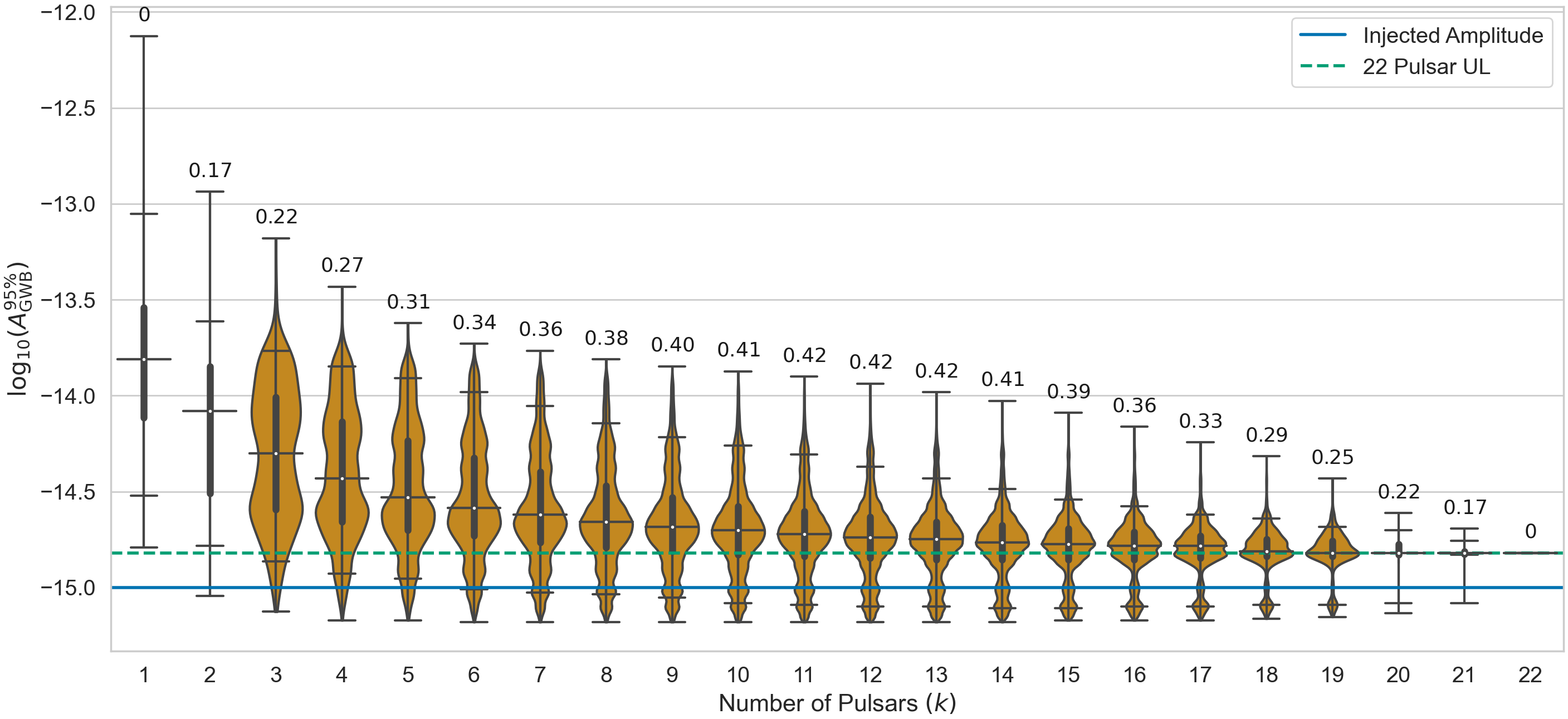}
\caption{Violin plot showing the distribution of upper limits given by combinations of $k$ pulsars for a single realization of the GWB.
The number of pulsars $k$ is given on the horizontal axis.
The minimum, 5\%, median, 95\%, and maximum values are given by horizontal lines on each violin.
The bold bar around the white dot showing the median value gives the 25\% to 75\% values.
Violins have been removed for values of $k$ that have fewer than 300 combinations.
This realization has a clear multi-modal structure in which one of the modes goes below the injected amplitude for subsets consisting of 15 or more pulsars. The number above each violin shows the number of combinations below the injected amplitude (blue, solid line) divided by the number of combinations below the 22 pulsar upper limit (green, dashed line) when using $k$ pulsars. When choosing pulsar combinations that give a lower upper limit than using the full set, this gives the probability of randomly selecting a combination that results in an upper limit below the injected amplitude.}
\label{fig:single}
\end{figure*}

\begin{figure*}[b]
\includegraphics[width=\textwidth]{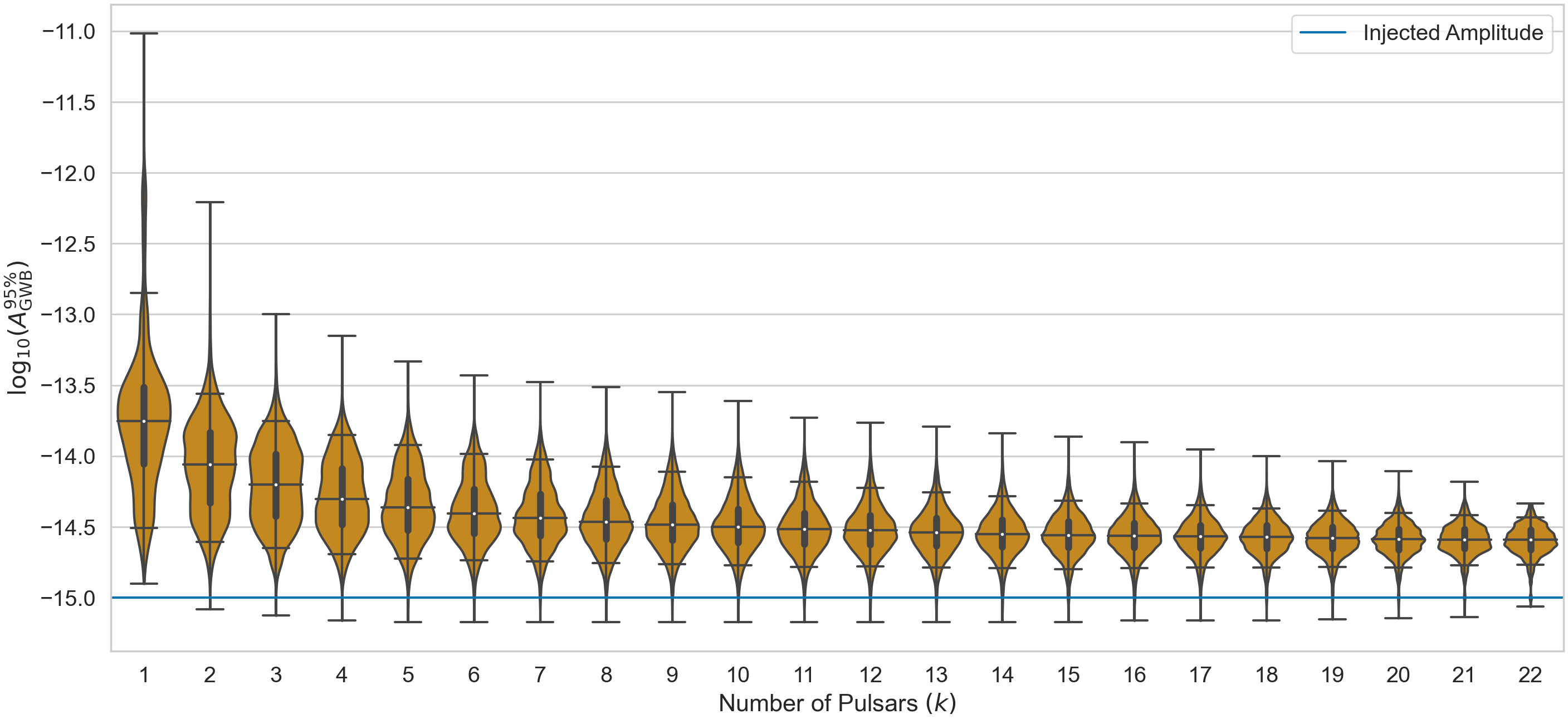}
\caption{Violin plot showing the distribution of upper limits given by combinations of $k$ pulsars averaged across 500 realizations. The number of pulsars $k$ is given on the horizontal axis.
The minimum, 5\%, median, 95\%, and maximum values are given by horizontal lines on each violin.
The bold vertical bars around the median value give the 25\% to 75\% values. The $\log_{10}A_\text{GWB}$ 95\% upper limit falls below the injected amplitude value in two to 53 (0.4\% to 10.6\%) of 500 realizations depending on the subset of pulsars used.}
\label{fig:tot}
\end{figure*}

Other than the combinations listed here, there are many other upper limit combinations that are possible.
The factorized PTA likelihood allowed us to compute the upper limits for all possible combinations of 22 pulsars -- a feat which would not be possible otherwise with current computers.
In Figure~\ref{fig:single}, we show the distributions of these upper limits for all $\binom{22}{k}$ combinations with a given $k$ on each violin.
This realization is worrisome because an entire mode of the multi-modal structure ends up below the injected amplitude.
The median upper limit decreases monotonically as we increase the number of pulsars and reaches its minimum value when using the full pulsar set.
As we increase the number of pulsars used, the spread of the distribution of possible upper limits decreases until we are left with only a single point when using the entire set of pulsars.
The number above each violin in Figure~\ref{fig:single} shows the number of combinations below the injected amplitude (blue, solid line) divided by the number of combinations below the 22 pulsar upper limit (green, dashed line) when using $k$ pulsars.
When choosing pulsar combinations that give a lower upper limit than using the full set, this gives the probability of randomly selecting a combination that results in an upper limit below the injected amplitude.
If we try to find an upper limit lower than when using the full data set in this realization, we risk ending up with an upper limit below the injected amplitude.

After investigating the combinations that end up below the injected amplitude, we find that there are some commonalities among these combinations.
Pulsars J1640+2224 and J1909-3744 are used in nearly every combination, while J1713+0747 is left out until all pulsars have been added.
However, we also find that the pulsars which are included and left out is realization dependent: another realization's combinations that fall below the injected amplitude do not necessarily include or exclude these particular pulsars.

Apart from being particularly influential on the upper limits (see Section \ref{sec:influencers}) these pulsars behave the same as all the other pulsars in our data set.
The recovered values of their intrinsic red noise amplitude and spectral indices return consistent with the injected values in every realization.
Further, we used simulations that do not include any extra astrophysical effects that may be mismodeled.
While removing J1640+2224 and J1909-3744 ``fixes'' this particular realization in the sense that the amplitude upper limit is no longer below the injected value, seven other realizations remain with upper limits below the injected amplitude.
Additionally, we cannot know beforehand that these pulsars cause problems without knowing the true value of the gravitational wave background amplitude.
Including the rest of the pulsars in the data set similarly pulls the upper limit back up to reasonable values in all but two realizations (0.4\%).

\subsection{All combinations of all realizations}
Figure~\ref{fig:tot} shows the distributions of upper limits averaged across all 500 realizations of the GWB.
The median upper limit again decreases monotonically as the number of pulsars increases and the range of upper limits decreases.
However, there are realizations that have upper limits that drop below the injected value with as few as two pulsars and as many as 22 pulsars.
When using the subset of pulsars that yields the lowest upper limit in all realizations, we find the $\log_{10}A_\text{GWB}$ upper limit below the injected value in 53 (10.6\%) out of 500 realizations.

Two realizations (0.4\%) remain below the injected value even when using the entire pulsar set.
In both cases, a single pulsar which strongly disfavors the GWB at and above its injected value dominates the upper limit with a marginalized $\log_{10}A_\text{GWB}$ posterior localized to values below the injected value.
Upon multiplying this pulsar's $\log_{10}A_\text{GWB}$ posterior by others, the other posteriors are forced to zero above the injected amplitude resulting in the overall upper limit falling below the injected amplitude.

\subsection{Single combination of all realizations}
\begin{figure}[b]
\centering
\includegraphics[width=\columnwidth]{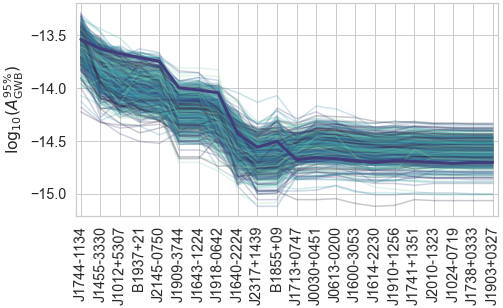}
\caption{Cumulative upper limits for a sequence of pulsars sorted by observation time from least to greatest. Each line corresponds to an individual realization (out of 500). The bold line represents one of these realizations. Most realizations in this sequence do not drop below the injected $\log_{10}A_{\mathrm{GWB}}$. Pulsars added in this combination have varied behavior between realizations: in some realizations, the pulsars increase the upper limit while in others they decrease the upper limit.}
\label{fig:time}
\end{figure}

Following the previous sections, we consider a single sequence of pulsars: from least observation time span to greatest.
This allows us to look at trends that exist across realizations from increasing the number of pulsars used in computing cumulative upper limits.
In Figure~\ref{fig:time} we show the cumulative upper limits obtained when adding pulsars in this order.
Each pulsar added decreases the upper limit on average until about 13 pulsars.
At this point, the upper limit saturates in this specific combination.
However, as shown in Figure~\ref{fig:uls}, this saturation does not happen for some pulsar combinations.

As can be seen in Figure~\ref{fig:time}, pulsars affect the upper limits differently between realizations.
For example, assuming pulsars are added in the same order, as they are here, J1713+0747 may lower the upper limit in one realization and increase the upper limit in the next.
However, some pulsars influence upper limits more than others.

\subsection{Influential pulsars}
\label{sec:influencers}

\begin{figure*}[b]
\includegraphics[width=\textwidth]{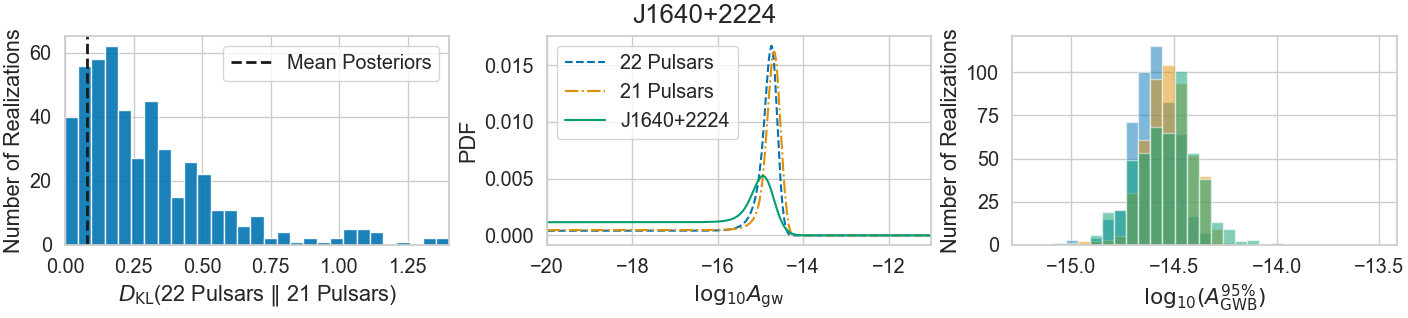}
\includegraphics[width=\textwidth]{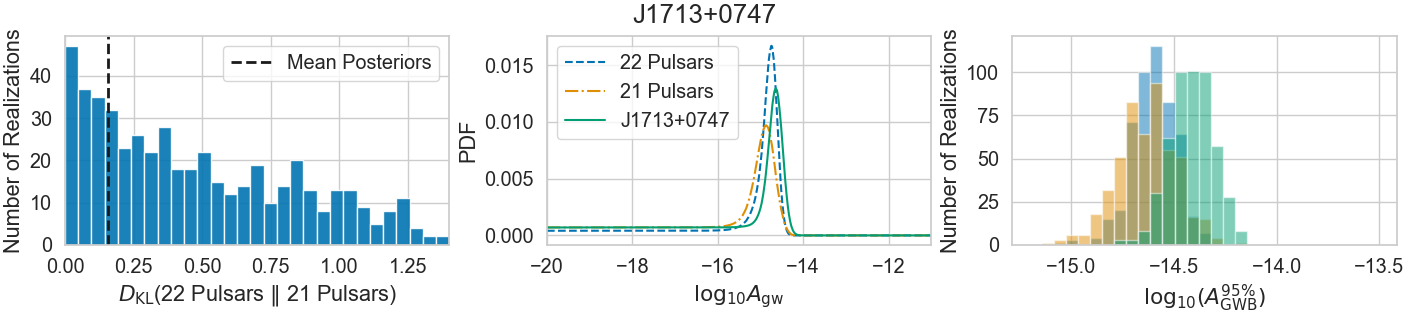}
\includegraphics[width=\textwidth]{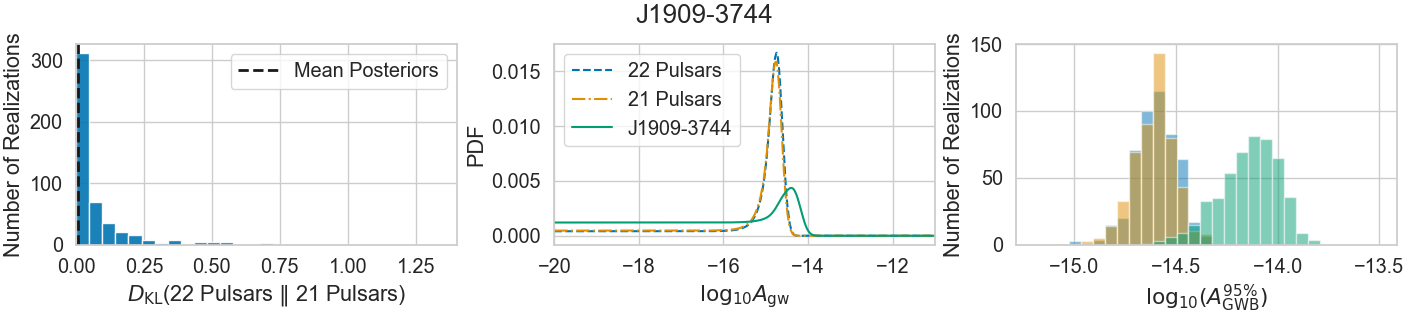}
\includegraphics[width=\textwidth]{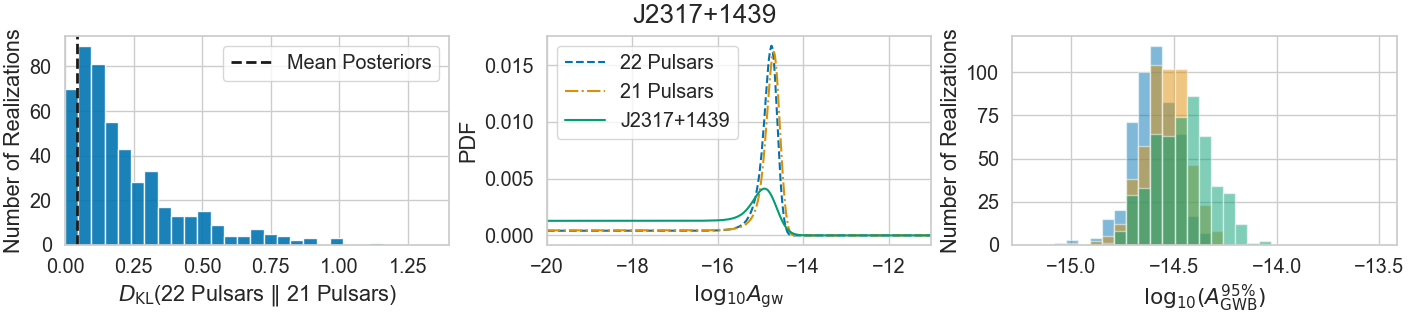}
\caption{Plots on the left column contain the KL divergence computed using the $\log_{10} A_{\mathrm{GWB}}$ posterior using all pulsars as the first argument and the $\log_{10} A_{\mathrm{GWB}}$ posterior using all but one pulsar as the second argument. The title of each subplot shows the pulsar that is removed from the second argument of the KL divergence. We cut out any pulsars that do not have $D_{\mathrm{KL}} > 0.5$. In the middle column, we plot the $\log_{10} A_{\mathrm{GWB}}$ posterior averaged over all realizations for the full 22 pulsar set, the 21 pulsar set, and the single pulsar which has been dropped. A vertical dashed line shows $D_{\mathrm{KL}}$ between the mean posteriors of 22 pulsars and 21 pulsars in the plots on the left column. On the right column, we plot the upper limits associated with each of the 500 realizations with colors that correspond to the legend in the middle column. Each divergence computed shows that even though these pulsars are often influential in the upper limits, there are realizations where the posterior with and without the pulsars are close.}
\label{fig:dropout}
\end{figure*}

In order to work out which pulsars are most influential for cumulative upper limits, we use the Kullback-Leibler (KL) divergence,
\begin{equation}
    \kldiv{P}{Q} = \sum_{x} P(x)\log\left(\frac{P(x)}{Q(x)}\right)
\end{equation}
as a measure of the difference between $P$ and $Q$, where we take $P(x)$ as the $\log_{10}A_{\mathrm{GWB}}$ posterior computed with all 22 pulsars in the data set, and we take $Q(x)$ as the $\log_{10}A_{\mathrm{GWB}}$ posterior computed while ``dropping out'' the pulsar whose influence we would like to check.
Importantly, $P$ remains the same for every pulsar that we are dropping out (although it will be slightly different between realizations).
Figure~\ref{fig:dropout} shows three columns for the pulsars that have $D_{\mathrm{KL}} > 0.5$ for any realization.
On the left column, it shows the KL divergence as described above. 
In the middle column, we plot the $\log_{10}A_{\mathrm{GWB}}$ posterior averaged over all realizations for the full data set, the full data set without one pulsar, and the pulsar that was dropped.
The $D_\mathrm{KL}$ value between the 22 pulsar mean posterior and the 21 pulsar mean posterior (both shown in the middle column) appears as a vertical line in the plots on the left column.
On the right column, we have histograms of upper limits for these same combinations for all 500 realizations.

As shown in Figure~\ref{fig:dropout}, J1640+2224 and J2317+1439 appear to slightly lower the mean posteriors and the upper limits once added to the set of pulsars used to compute upper limits.
J1713+0747 tends to increase the mean posterior and upper limits, and J1909-3744 does not affect the mean posterior or upper limits significantly in either direction.
In every pulsar on these plots, there are realizations in which adding the pulsars does not significantly change the posterior, and we see this manifest as a $D_\mathrm{KL} \approx 0$.
Pulsars not shown in these plots have smaller KL divergences between the full pulsar set and with one pulsar removed.
This does not mean that we should drop these pulsars when computing upper limits, but that they show less influence on the overall upper limit once 21 pulsars have been included.

This analysis explains the results of the realization in Section~\ref{sec:single_combos}. The two pulsars that were included in most of the combinations below the injected amplitude, J1640+2224 and J1909-3744, appear in this list of our most influential pulsars in these simulated data sets.
Further, J1640+2224 tends to move the upper limit in a downward direction.
J1713+0747, in contrast, tends to move the upper limit in an upward direction, and therefore it is left out until the last few pulsars.

\subsection{Bias}
Computing all possible combinations of upper limits, we find that many realizations have combinations that result in upper limits below the injected amplitude value of the GWB.
Out of 500 realizations, 53 (10.6\%) realizations have at least one combination that gives a 95\% upper limit below the injected value.
This number drops to two (0.4\%) realizations when using the full data set for every realization.

\begin{figure}[h]
\includegraphics[width=\columnwidth]{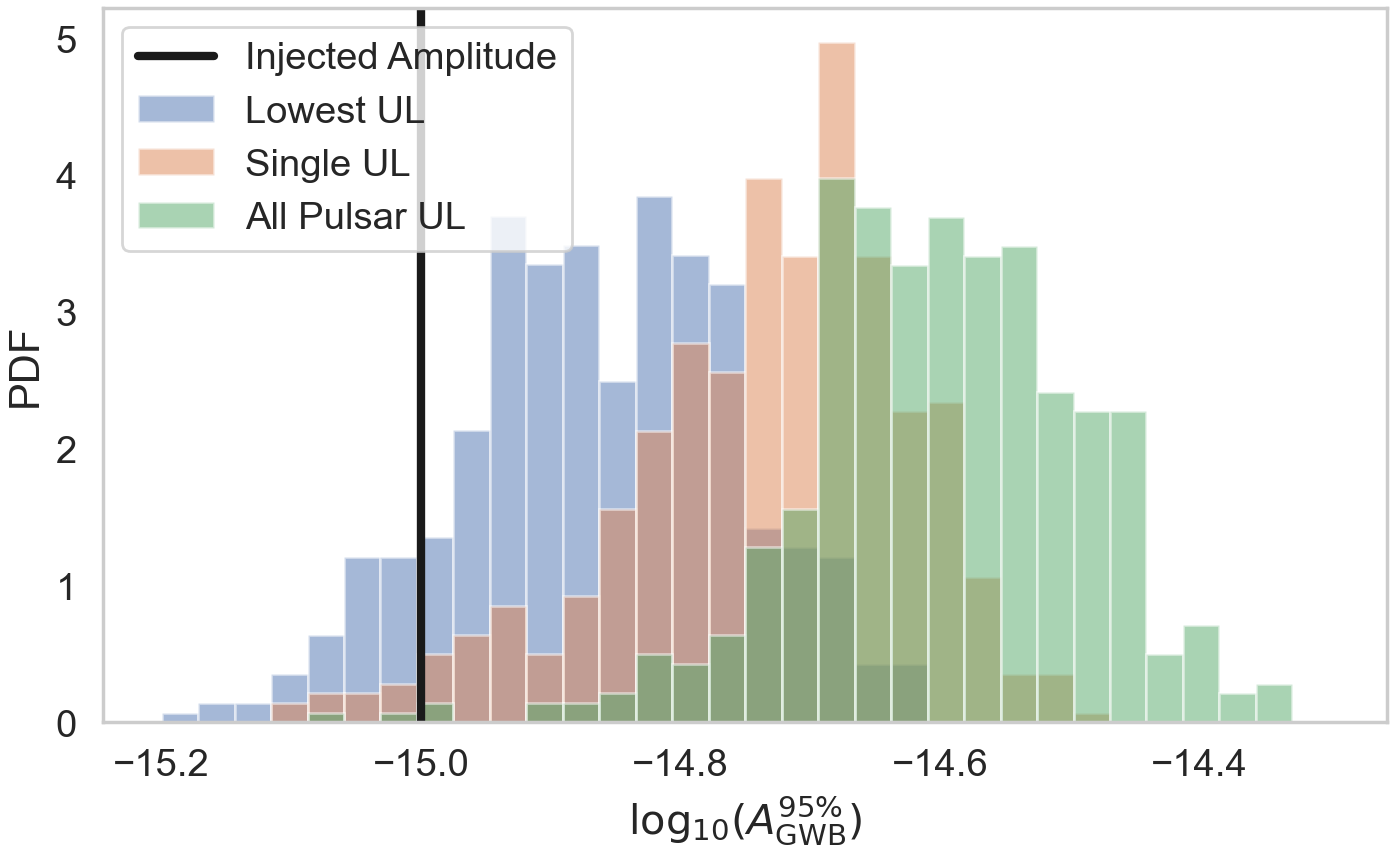}
\caption{The upper limits obtained for all 500 realizations, where the upper limits have been found using three different methods.
In one method, we computed the upper limit using all 22 pulsars (green histogram on the right).
In another method, we ranked the pulsars based on their single-pulsar upper limits on the GWB, and then combined those pulsars one by one until adding another pulsar caused the upper limit to increase.
We took this minimum upper limit as the true upper limit (orange histogram in the middle).
In the final method, we looked at all possible combinations of pulsars and chose the lowest possible upper limit that could be obtained (blue histogram on the left).
Upper limits obtained using either the second or third method tend to be lower than those obtained using all 22 pulsars resulting in a systematic shift toward lower amplitudes for these histograms (blue on the left and orange in the middle).
The blue histogram on the left has 53 (10.6\%) realizations below the injected amplitude (vertical, black line), the orange histogram in the middle has 12 (2.4\%) realizations below the injected amplitude, and the green histogram on the right has two (0.4\%) realizations below the injected amplitude.}
\label{fig:bias}
\end{figure}

Figure~\ref{fig:bias} shows the upper limits obtained for all 500 realizations, where the upper limits have been found using three different possible methods.
In one method, we computed the upper limit using all 22 pulsars (green histogram on the right).
In another method, we ranked the pulsars based on their single-pulsar upper limits on the GWB, and then combined those pulsars one by one until adding another pulsar caused the upper limit to increase.
We took this minimum upper limit as the true upper limit (orange histogram in the middle).
In the final method, we looked at all possible combinations of pulsars and chose the lowest possible upper limit that could be obtained (blue histogram on the left).

Note that when using the second or third method, we necessarily end up using either all 22 pulsars or using some subset of them, and the computed upper limits must be either equivalent to the upper limit obtained using all 22, or it must be smaller, which is why the blue (left) and orange (middle) histograms are shifted to lower values relative to the green histogram.
As shown Figure~\ref{fig:bias}, upper limits obtained using either the second or third method tend to be lower than those obtained using all 22 pulsars; furthermore, we find 12 (2.4\%) realizations of the orange (middle) histogram and 53 (10.6\%) realizations of the blue (left) histogram out of 500 have an upper limit that falls below the injected amplitude.
These results demonstrate that choosing a subset of pulsars in order to obtain the lowest possible upper limit results in a biased measurement.

\section{Discussion and Conclusion}
\label{sec:conclusion}
In this paper, we use simulated PTA data to study how the choice of which pulsars to include in GWB analyses can bias upper limits on the GWB. 
By factorizing the PTA likelihood \citep{taylor_parallelized_2022}, we are able to compute all possible combinations of upper limits for each realization of the GWB.
This method limits us to an auto-correlation only analysis.
However, we find that including cross-correlations does not change the upper limits in the cases that the upper limit falls below the injected value.
In every realization of 500, we find that the median and spread of the distribution of upper limits decrease monotonically as the number of pulsars used increases.
In some realizations, the probability of finding a value below the injected amplitude is significant when picking combinations that give upper limits below those returned by using all 22 pulsars.
When using all pulsars to set the upper limit, we find that the upper limit is below the injected value in just two of the 500 realizations.

By investigating the sequences of upper limits of different combinations of pulsars, we have shown that we can bias our upper limit to lower $\log_{10}A_\text{GWB}$ values by using a subset of pulsars.
In 53 (10.6\%) of 500 realizations, the upper limit falls below the injected amplitude when choosing the minimum value in the lowest upper limit combination sequence.
While this is the maximum bias to lower values of $\log_{10}A_\text{GWB}$ that we can find, it is far from the only set of combinations that are biased toward lower values.
Picking the lowest upper limit of any sequence of upper limits given by a combination of pulsars will \textit{always} result in either the full set of pulsars being used or result in a distribution of upper limits from the 500 realizations shifted toward lower $\log_{10}A_\text{GWB}$ values.

Multiple pulsar timing array experiments have recently published results reporting the detection of a common stochastic process whose amplitude is in tension with previously published upper limits on the amplitude of such a process. This work helps to explain one possible reason for this discrepancy. 
Earlier published work used significantly fewer pulsars compared to the number being used in the most recent papers, and as shown in this paper, using a small number of pulsars to set upper limits can lead to bias and can even result in upper limits that are \textit{lower} than the true amplitude. 
The range of possible upper limits decreases as we increase the number of pulsars, 
and therefore 
using as many pulsars as we can reduces the probability that the upper limit that we get is below the actual value of the GWB.
Furthermore, using more pulsars has the added benefit of producing a finer grid of angular separations of pulsar pairs, increasing our sensitivity to the cross-correlations that are characteristic of the GWB. 
In order to avoid bias and improve detection prospects, the best strategy for PTAs is to include as many MSPs as possible.

\section*{acknowledgements}
We thank Jeff Hazboun for helpful suggestions early on during this work.
Joe Simon provided code to combine posteriors from individual pulsar analyses into posteriors from a joint analysis.
We also thank Michele Vallisneri for valuable comments.
Finally, we thank the anonymous reviewer, whose suggestions led to new interesting plots and discussion.

This work was supported by NSF Physics Frontier Center Grant 1430284. 
ADJ and SJV were supported by UWM Discovery and Innovation Grant 101X410.
SJV was supported by NSF PHY-2011772.
SRT acknowledges support from NSF AST-200793, PHY-2020265, and PHY-2146016.
SRT also acknowledges support from Vanderbilt University College of Arts \& Science Dean's Faculty Fellowship program.

\bibliography{bib}

\end{document}